\documentclass[useAMS,usenatbib,usegraphicx,a4paper]{mn2e}

\usepackage{hyperref}
\hypersetup{dvips,pdfauthor={Federico Garcia},
            pdftitle={The cooling phase of Type I X-ray bursts observed with RXTE in 4U 1820-30}}
\title[The cooling phase of Type I X-ray bursts observed with {\it RXTE} in 4U 1820--30]{The cooling phase of Type I X-ray bursts observed with {\it RXTE} in 4U 1820--30 does not follow the canonical $F \propto T^4$ relation}
\author[F. Garc\'{\i}a et al.]
{Federico Garc\'{\i}a$^{1}$~\thanks{Fellow of CONICET; e-mail: fgarcia@iar-conicet.gov.ar}, Guobao Zhang$^{2}$ and Mariano M\'{e}ndez$^{2}$\\
$^{1}$Instituto Argentino de Radioastronom\'{\i}a (CCT La Plata, CONICET), C.C.5, (1894) Villa Elisa, Buenos Aires, Argentina\\
$^{2}$Kapteyn Astronomical Institute, University of Groningen, PO Box 800, 9700 AV Groningen, the Netherlands}
\begin{document}

\date{Accepted ; Received }

\pagerange{\pageref{firstpage}--\pageref{lastpage}} \pubyear{2012}

\maketitle

\label{firstpage}

\begin{abstract}
We analysed the complete set of bursts from the neutron-star low-mass X-ray binary 4U 1820--30 detected with the {\it Rossi X-ray Timing Explorer (RXTE)}. We found that all are photospheric radius expansion bursts, and have similar duration, peak flux and fluence. From the analysis of time-resolved spectra during the cooling phase of the bursts, we found that the relation between the bolometric flux and the temperature is very different from the canonical $F \propto T^4$ relation that is expected if the apparent emitting area on the surface of the neutron star remains constant. The flux-temperature relation can be fitted using a broken power law, with indices $\nu_1$=2.0$\pm$0.3 and $\nu_2$=5.72$\pm$0.06. The departure from the $F \propto T^4$ relation during the cooling phase of the X-ray bursts in 4U 1820--30 could be due to changes in the emitting area of the neutron star while the atmosphere cools-down, variations in the colour-correction factor due to chemical evolution, or the presence of a source of heat, e.g. residual hydrogen nuclear burning, playing an important role when the burst emission ceases.
\end{abstract}


\section{Introduction}

Accreting neutron stars (NSs) in low-mass X-ray binaries (LMXBs) show episodes of unstable thermonuclear burning of hydrogen and helium on their surfaces \citep[e.g.][]{lewin1993,strohmayer2006,galloway2008}. These episodes are called Type I X-ray bursts. Although the vast majority of the energy released in LMXBs originates in the accretion process, capable of radiating $\sim$300~MeV per accreted baryon, the unstable fusion process ($\la$5~MeV per nucleon) during an X-ray burst can increase the luminosity of these objects by a factor of $\sim$10 in about 0.5--5~s. After that fast rise, the X-ray flux decreases again exponentially within 10--100~s. The typical total energy emitted during an X-ray burst is $\sim$10$^{39}$~erg. Some of these bursts are strong enough to reach the Eddington luminosity. In those cases, the radiation pressure is high enough to trigger the expansion of the outer layers of the NS atmosphere, in a so-called Photospheric Radius Expansion (PRE) burst \citep{basinska1984}. The hypothesis that the bursts originate at the NS surface is supported by the fact that the inferred emission area during the bursts matches the expected surface area of a standard NS (i.e. $\sim$400~km$^2$), if a symmetric cooling phase is assumed \citep{fryxell1982,bildsten1995,spitkovsky2002,strohmayer2006}. This emitting area can be estimated from the time-resolved spectral analysis during the cooling phase of the bursts.

The X-ray bursting LMXB 4U 1820--30 is an ultracompact X-ray binary, with an orbital period of 11.4~min \citep{stella1987}, consisting of a NS and a hydrogen-exhausted white-dwarf of 0.06--0.08~M$_{\odot}$\citep{rappaport1987}. 4U 1820--30 belongs to the globular cluster NGC 6624, which is located at a distance of 7.6$\pm$0.4~kpc \citep{kuulkers2003}. Thermonuclear X-ray bursts in this LMXB were discovered by \cite{grindlay1976}. The source also displays kilohertz quasi-periodic oscillations \citep{smale1997,zhang1998}. Until 2007 June, five Type I X-ray bursts --all PRE-- were detected by the {\it Rossi X-ray Timing Explorer}, {\it RXTE} \citep{galloway2008}. From time-resolved spectra of three of these five bursts, \cite{guver2010} found that, during the cooling phase, the bolometric flux was proportional to the temperature to the fourth power, and from this they estimated the mass and radius of the NS in 4U 1820--30. \cite{strohmayer2002} also discovered a superburst in this source, presumably due to the burning of a deep carbon layer. In order to explain the characteristics of these X-ray bursts, \cite{cumming2003} developed models of a NS accreting He-rich material. 

In this paper we study the complete sample of X-ray bursts, 16 in total, in 4U 1820--30 detected with {\it RXTE}. We explore the properties of these bursts and we focus on their behaviour during the cooling phase. Finally, we compare our results with those from \cite*{zhang2011} on the cooling phase of 4U 1636--56, and with recent NS atmosphere models developed by \cite{suleimanov2011}.

\section[]{Observations and data analysis}

We analysed all archival data from the {\it RXTE} Proportional Counter Array (PCA) of 4U 1820--30. These observations cover the period from May 1996 to April 2011. The PCA detector consists of five collimated proportional counter units (PCUs) sensitive in the 2--60~keV energy range. In order to detect the X-ray bursts in the dataset, we generated 1-s time-resolution light-curves from the Standard-1 mode (data packed in $0.125$-s time-resolution bins without energy resolution) and we analysed each light-curve to search for three consecutive points that were \mbox{8-$\sigma$ above the average count rate} of the observation. We have chosen this value to distinguish real bursts from fluctuations in the persistent emission. We confirmed each burst detection by inspecting all the Standard-1 light-curves in the whole dataset by eye. We used Standard-2 mode data (16-s time-resolution and 129 energy channels) to produce X-ray colours of the source, following the procedure described in \cite{zhang2009}. We selected four energy bands: $A=$ 2.0--3.5~keV, $B=$ 3.5--6~keV, $C=$ 6.0--9.7~keV and $D=$ 9.7--16.0~keV interpolating in channel space, and we defined soft and hard colours as count rate ratios SC $=B/A$ and HC $=D/C$, respectively. We also defined the X-ray intensity, $I_{\rm X}$, as the 2.0--16.0~keV count rate. The count rates in each energy band were normalized using the PCA data of the Crab source obtained close in time to our observations. 

\begin{table*}
 \centering
  \begin{minipage}{120mm}
  \caption{X-ray bursts in 4U 1820--30 detected with {\it RXTE}.}
  \begin{tabular}{rlcccccc}
  \hline
& Observation ID & Starting time & Duration & SC $^{*}$ & HC $^{*}$ & $I_{\rm X}$ $^{*}$\\
&			     & [MJD] &    [s]   & [Crab]& [Crab]& [Crab]   \\
 \hline
$\dagger$ & 20075-01-05-00 & 50570.73116038 & $\sim$26 & 1.11(1) & 0.91(1) & 0.1263(3) \\
& 30057-01-04-08G & 51430.0461502 & $>$3000 $^{**}$ & 1.08(4) & 0.80(4) & 0.112(1) \\
$\dagger$ & 40017-01-24-00 & 52794.73816848 & $\sim$28 & 1.19(2) & 1.00(1) & 0.0894(4) \\
$\dagger$ & 70030-03-04-01 & 52802.07560482 & $\sim$26 & 1.15(2) & 0.98(1) & 0.0914(4) \\
$\dagger$ & 70030-03-05-01 & 52805.89569741 & $\sim$25 & 1.19(1) & 1.01(1) & 0.1238(4) \\
$\dagger$ & 90027-01-03-05 & 53277.43859093 & $\sim$23 & 1.09(1) & 0.89(1) & 0.1317(5) \\
& 94090-01-01-02 & 54948.82125875 & $\sim$28 & 1.15(1) & 0.98(1) & 0.1130(4) \\
& 94090-01-01-05 & 54950.70282704 & $\sim$30 & 1.22(2) & 1.01(1) & 0.1012(4) \\
& 94090-01-02-03 & 54956.77474255 & $\sim$25 & 1.26(2) & 1.09(2) & 0.0949(4) \\
& 94090-01-02-02 & 54958.73999718 & $\sim$31 & 1.27(2) & 1.10(1) & 0.0912(4) \\
& 94090-01-04-00 & 54978.32150760 & $\sim$30 & 1.33(2) & 1.11(2) & 0.1052(5) \\
& 94090-01-04-01 & 54978.49491038 & $\sim$29 & 1.30(2) & 1.14(1) & 0.1048(5) \\
& 94090-01-05-00 & 54981.18730621 & $\sim$26 & 1.29(2) & 1.12(1) & 0.1130(5) \\
& 94090-02-01-00 & 54994.53419278 & $\sim$21 & 1.18(1) & 1.02(1) & 0.1855(6) \\
& 94090-02-01-00 & 54994.61302380 & $\sim$22 & 1.19(1) & 0.99(1) & 0.1905(7) \\
& 96090-01-01-00G & 55624.88068005 & $\sim$25 & 1.22(2) & 1.01(1) & 0.0940(4) \\
& 96090-01-01-020 & 55626.77361987 & $\sim$21 & 1.11(1) & 0.83(1) & 0.1707(6) \\
\hline
\end{tabular}
\\ $^{*}$ Soft/Hard colours and intensity of the persistent emission just before the X-ray burst, in Crab units.
\\ $^{**}$ Superburst, see \cite{strohmayer2002}.
\\ $\dagger$ X-ray bursts analysed by \citet{guver2010}.
\label{bursts_table}
\end{minipage}
\end{table*}

\begin{figure}
	\centering
 \includegraphics[width=8.4cm]{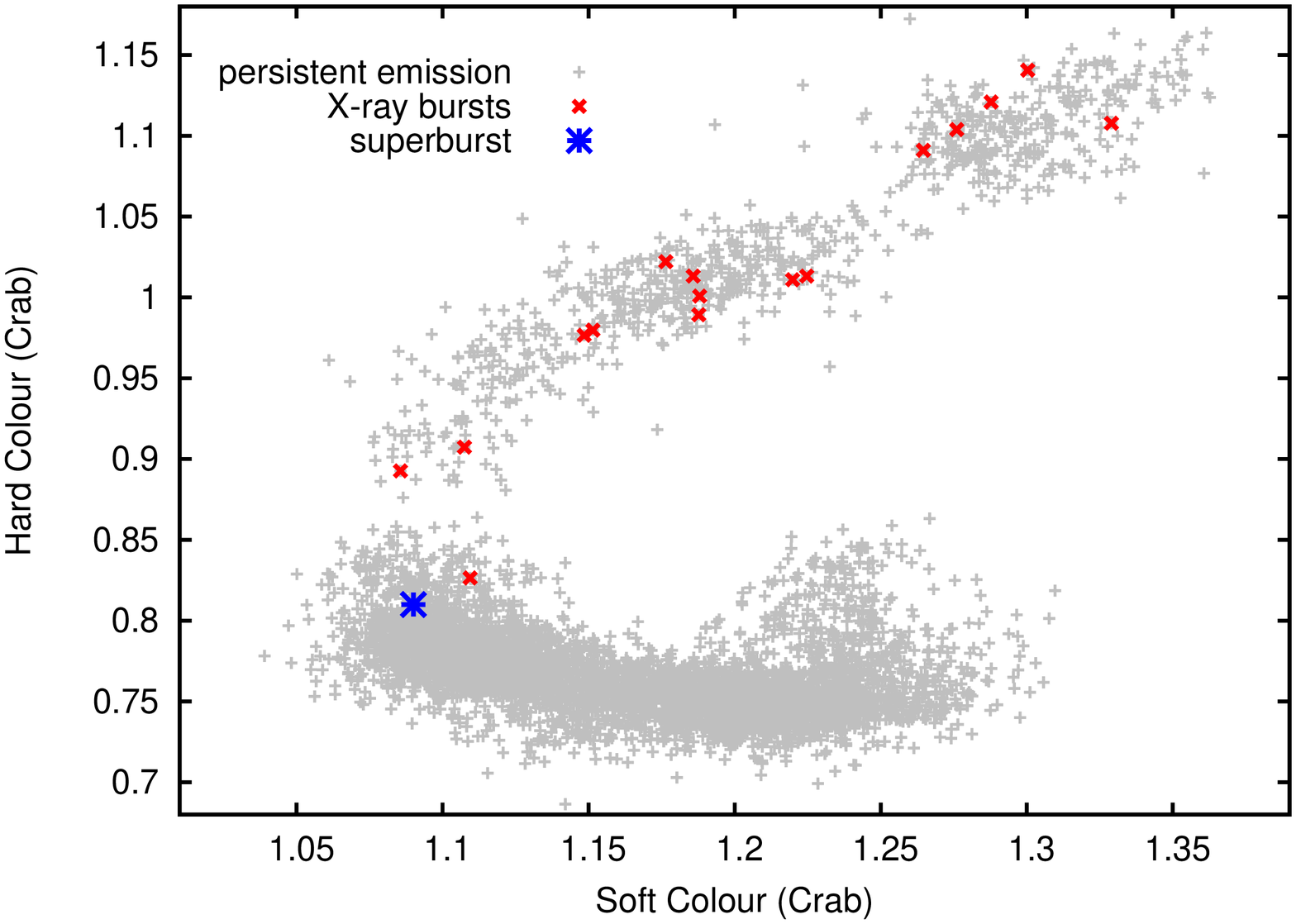}
 \caption{Colour-colour diagram of all {\it RXTE} observations of \mbox{4U 1820--30}. Grey points correspond to the whole set of observations with 256-s time resolution. The colour of the persistent emission previous to the 16 X-ray bursts is indicated with red crosses. The colour of the persistent emission previous to the superburst is indicated with a blue star. Soft and hard colours are in Crab units.}
 \label{CD}
 \vspace{1cm}
	\centering
   \includegraphics[width=8.4cm]{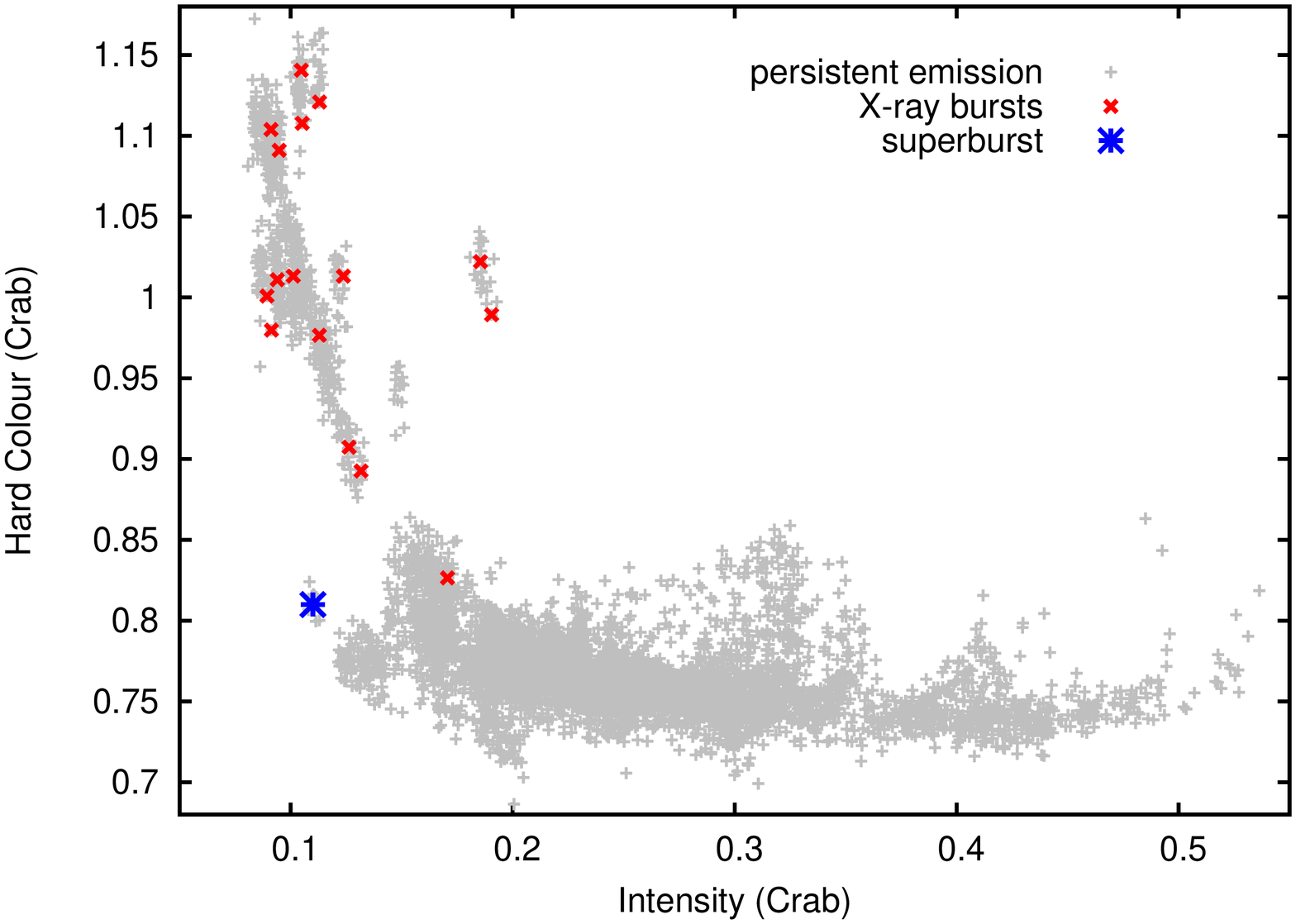}
  \caption{Hardness-intensity diagram of all {\it RXTE} observations of 4U 1820--30. The symbols are the same as in Figure~\ref{CD}. Intensity and hard colour are in Crab units.}
  \label{HID}
  
\end{figure}

To study in detail each detected X-ray burst, we used the PCA Event mode data, E\_125us\_64M\_0\_1s, where data is binned in 64 energy channels in the  2--60~keV energy range, with 125-$\mu$s time-resolution. For the time-resolved analysis of the bursts we extracted spectra using the 64 channels in time-intervals going from, respectively, 0.125 s at the peak to 1-2 s at the tail of the burst. We corrected each spectrum for dead-time using the methods supplied by the {\it RXTE} team\footnote{http://heasarc.gsfc.nasa.gov/docs/xte/whatsnew/deadtime.html} and we generated response matrix files for the PCUs active during each observation containing bursts. We fitted the data using {\sc XSPEC} version 12.7.0 \citep{arnaud1996}, assuming 0.5 per cent systematic error, as suggested by the {\it RXTE} calibration team\footnote{http://www.universe.nasa.gov/xrays/programs/rxte/pca/\\doc/rmf/pcarmf-11.7/}. For all time-resolved spectra (250 in total) we fitted the data from 2.5 to 20 keV. An artifact in the binning scheme of the Event mode data leads to zero counts in one channel per PCU unit \citep{jahoda1996}. This is normally accounted for in the response matrix files. We found, however, that if only one PCU was on during an observation, that single channel with zero counts disrupted the fits. In our case, this happened in a handful of spectra in which only PCU2 was on, and channel 11 had zero counts. For this reason we ignored channel 11 ($E\sim4$~keV) in all our spectra. We ended up with 25 channels in the 2.5--20~keV range for each spectrum.
We extracted $\sim$100~s spectrum from the persistent emission before and after each burst and then used this, independently, as background for the burst spectra, following the standard procedure in X-ray bursts analysis \citep[e.g. ][]{kuulkers2002}. We fitted the spectra using a blackbody model ({\it bbodyrad}) times a component that accounts for the interstellar absorption towards the source ({\it tbabs}) fixing the hydrogen column density, $N_{\rm H}$. The model for the interstellar absorption uses the abundances of \cite{wilms2000} and the cross sections of \cite{verner1996}. The spectral model consists of two free parameters: the blackbody temperature, $kT_{\rm bb}$, in keV units (where $k$ is the Boltzmann constant) and the blackbody normalization, $N_{\rm bb}=(R_{\rm bb}/d_{10})^2$, where $R_{\rm bb}$ is the blackbody radius in km and $d_{10}$ is the distance to the source in units of 10~kpc. These parameters allow us to estimate the evolution of the bolometric flux during the bursts using, e.g., eq. (3) of \cite{galloway2008},
$F = 0.001076 N_{\rm bb} (kT_{\rm bb})^4 \hspace{0.3cm}10^{-8}{\rm ~erg~cm}^{-2}{\rm ~s}^{-2}$.

This procedure fails if the blackbody observed during the burst comes from the same blackbody that produces the persistent emission before the burst, because the difference between two blackbody spectra is not a blackbody \citep{vanparadijs1986}. However, in practice, the deviations from a Planckian distribution become important only if the net emission is weak. Therefore, this issue may be important only at the very beginning or at the tail of the bursts.

To test whether using the parameters of the fitting of the net spectrum instead of the total spectrum --which includes modelling the persistent emission during the burst and subtracting the instrumental background-- could affect our final results, we proceeded as follows: we fitted the persistent spectrum before each burst using an absorbed blackbody plus a power-law model. Then, for the bursting time, we fixed the power-law index and normalization to the best-fitting values obtained for the persistent emission and we fitted only the blackbody parameters. The results from this procedure are consistent with those we obtained by fitting the net spectrum within errors, either using the persistent emission before or after each burst as background. To study whether possible variations in the persistent emission during the bursting time could change our results, we also fitted the spectra leaving free the power-law component. The results of this analysis are fully consistent within the errors with those found by fitting the net spectra. We also evaluated the contribution of the persistent emission to the net spectra during bursting time; we found that, for every spectra, the blackbody contributes at least 25 per cent of the total emission in the 2.5--20~keV energy band and in 75 per cent of the cases the blackbody contribution in this energy range is larger than 50 per cent. In the rest of this work we show the results from the standard procedure. 

In all spectra the number of counts per channel in the source and in the background is high enough to justify the use of minimum $\chi^2$ \citep{pearson1900} to find the best-fitting parameters. Therefore, contrary to \cite{guver2010}, we did not rebin the spectra in energy (we note also that the original binning scheme of the event-mode data that we used in the spectral analysis oversamples the intrinsic spectral resolution of the PCA by a factor of 1 to 3, depending on energy, and hence rebinning is neither required nor desirable). For the sake of comparison, we also fitted all the spectra using the C statistics defined by \cite{cash1979}, which is the {\sc XSPEC} implementation of the maximum-likelihood estimate for the case of data following Poisson statistics. We found consistent results with both methods, and therefore here we report the results from the minimum $\chi^2$ method. In the plots we show the \mbox{1-$\sigma$} errors ($\Delta \chi^2 = 1$). 

\section[]{Results}

\subsection{X-ray burst detection and colour analysis}

We identified 16 short bursts (20--30~s), and the presence of one superburst ($>$3000~s), previously reported by \cite{strohmayer2002} in all the {\it RXTE} archival data. In Table~\ref{bursts_table} we summarize the main properties of the X-ray bursts. The starting time of the bursts was defined when a continuous increase in the PCA count rate persisted for at least 0.25~s within each detected burst (see Section 2), while the ending time was defined as the time when the burst count rate returned to within \mbox{3-$\sigma$} of the flux of the persistent emission before the onset of the burst. We also show in the table the burst duration --the difference between ending and starting times-- and the soft and hard colours and the intensity of the persistent emission before each burst, derived from the Standard-2 mode data of the PCA.

Figures~\ref{CD} and \ref{HID}, respectively, show the position of the source in the CD and HID before each burst. No bursts were detected for a persistent ${\rm HC}<0.8$ or for $I_{\rm X}>0.2$~Crab, where the source is bright and soft and the accretion rate is likely high \citep{hasinger1989}. 

\begin{figure}
\vspace{0.2cm}
\centering
 \includegraphics[width=7.6cm]{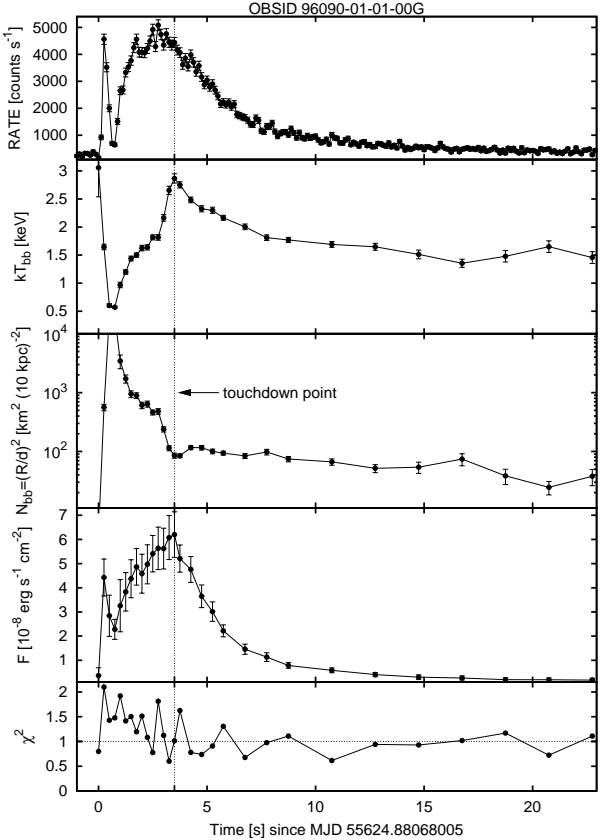}
  \caption{The X-ray burst in 4U 1820--30 detected in ObsID 96090-01-01-00G. The panels from top to bottom correspond to the Standard-1 mode (0.125-s time-resolution) X-ray light-curve (starting 1~s before the burst detection), the blackbody temperature, $kT_{\rm bb}$, normalization, $N_{\rm bb}$, bolometric flux, $F$, and reduced $\chi^2$ for 23 degrees of freedom of the best-fit to the time-resolved spectra (starting from the onset of the burst), with 1-$\sigma$ error-bars. The vertical dotted line indicates the touchdown point, which defines the starting of the cooling phase of the PRE burst. }
  \label{fitting}
\end{figure}    

\subsection{Time-resolved spectral analysis of the X-ray bursts}

In Figure~\ref{fitting} we show the results of the fits to one of the X-ray bursts detected in 4U 1820--30. Because all the bursts in our dataset are very similar, we only show one of them as an example to illustrate the observed burst characteristics and the analysis process. In the top panel we present the Standard-1 high time-resolution light-curve, starting 1~s before the onset of the burst. When the burst ignites, the light-curve shows a very high and narrow peak lasting for 1--2~s. After that feature, the count rate starts to increase again reaching a maximum of $\sim$5000~counts~s$^{-1}$ in 2--3~s, and finally it decays exponentially until the persistent count rate is recovered in 15--25~s. In the next four panels we present the evolution of the best-fitting parameters $kT_{\rm bb}$ and $N_{\rm bb}$, the bolometric flux derived from the blackbody model and the reduced $\chi^2$ residuals of the best-fitting model for 23 degrees of freedom, respectively. 
The distribution of the best-fitting $\chi^2$ of the 250 spectra is consistent with the expected $\chi^2$ distribution. A Kolmogorov-Smirnov test gives a probability of 0.09, which is not enough to disprove the hypothesis that the best-fitting $\chi^2$ are drawn from the expected $\chi^2$ distribution for 23 degrees of freedom.
During the spectral fitting, we considered two different values for the hydrogen column density, following \cite{costantini2012} and \cite{guver2010}: $N_{\rm H}=0.16$ and $0.25\times10^{22}$~cm$^{-2}$, respectively. The best-fitting spectral parameters obtained for these column densities are fully consistent with each other, because in the 2.5--20~keV energy band, absorption is not very significant. From now on, we show results obtained assuming $N_{\rm H}=0.25\times10^{22}$~cm$^{-2}$.

At the beginning of the burst the blackbody temperature, $kT_{\rm bb}$, starts at $\sim$3~keV and decreases rapidly, while the blackbody normalization, $N_{\rm bb}$, which is proportional to the inferred emitting area, increases very fast. Then, $kT_{\rm bb}$ starts to increase again as $N_{\rm bb}$ decreases, until the touchdown point is reached. At this point, the $N_{\rm bb}$ is minimum and the temperature is maximum. Thus, we define this time as the start of the cooling phase. This common behaviour seen in all bursts indicates that they are all PRE. The high and narrow peak observed in the Standard-1 light curves at the start of the burst \citep{keek2012} is due to the fact that the temperature decreases enough to move the Planckian flux distribution to the soft band, outside the energy range where the PCA is sensitive \citep{tawara1984,intzand2010}. 

In order to study the properties of the whole sample of bursts, we analysed the behaviour of four parameters inferred from the X-ray analysis with respect to the position of the bursts in the CD. These parameters are: the temperature, $kT_{\rm peak}$, and flux, $F_{\rm peak}$, at the touchdown point, the fluence, $B$, defined as the total energy per unit area emitted during the burst, $B=\int F(t)dt$, and the decay time scale, $\tau$, obtained from the fit of an exponential function to the flux during the cooling phase of the burst. We divided the bursts in two groups, based on the hard colour of the persistent emission at the time of the burst: ${\rm HC}>1.05$ and ${\rm HC}<1.05$. The statistical analysis of both subsamples shows that both populations are indistinguishable through these parameters \citep[see also][]{intzand2012}. 

\begin{figure}
\centering
 \includegraphics[width=8.4cm]{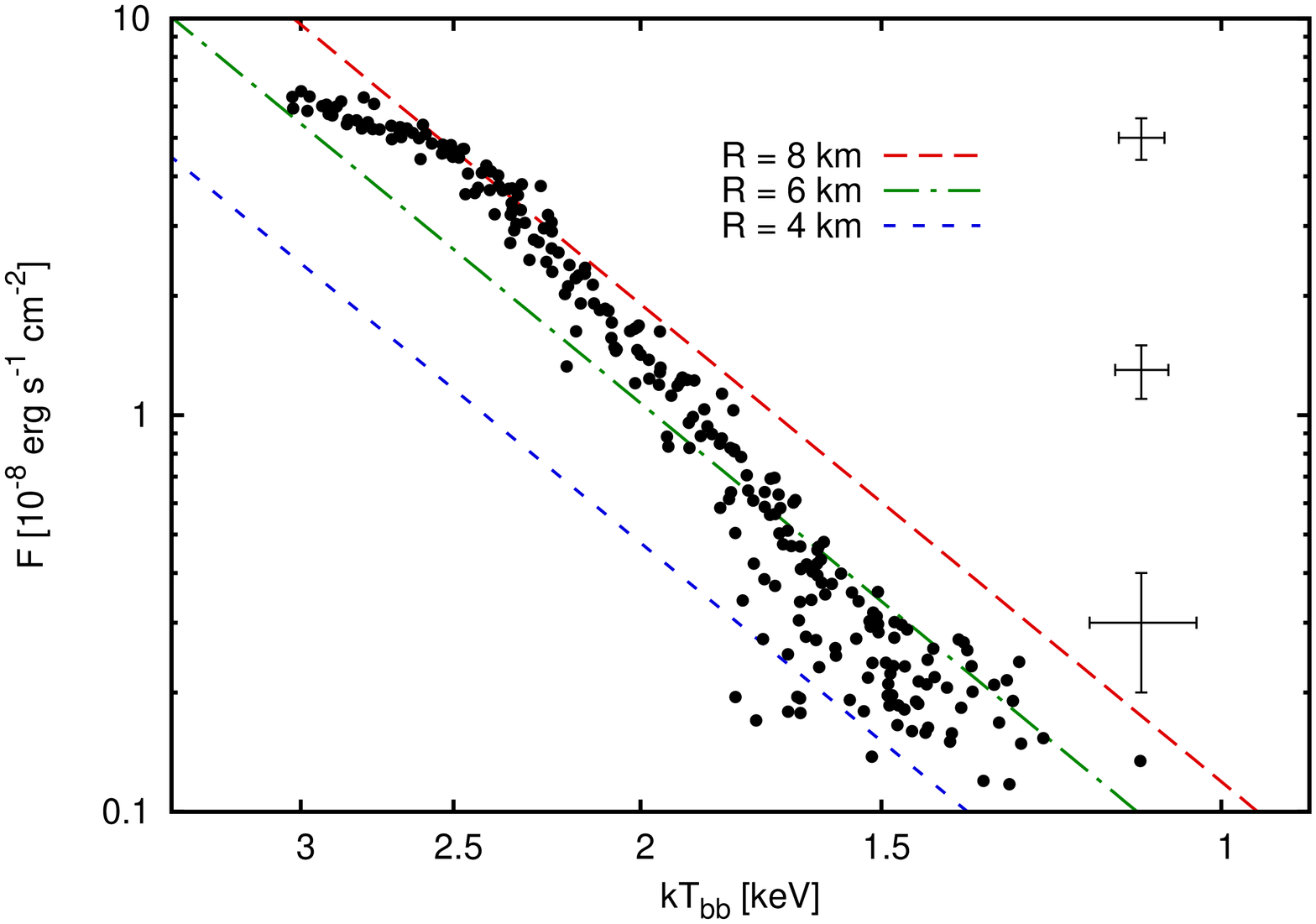}
\caption{Merged plot of the cooling phase of the 16 X-ray bursts of 4U 1820--30 detected with {\it RXTE}. The three diagonal lines are lines of constant radius, corresponding, respectively, to three blackbodies of $R=4$, 6 and 8~km, assuming a distance of 7.6~kpc to the source. We show the typical error bars for three different bolometric flux levels at the right side of the plot.}
\label{cooling}
\vspace{0.5cm}
\centering
 \includegraphics[width=8.4cm]{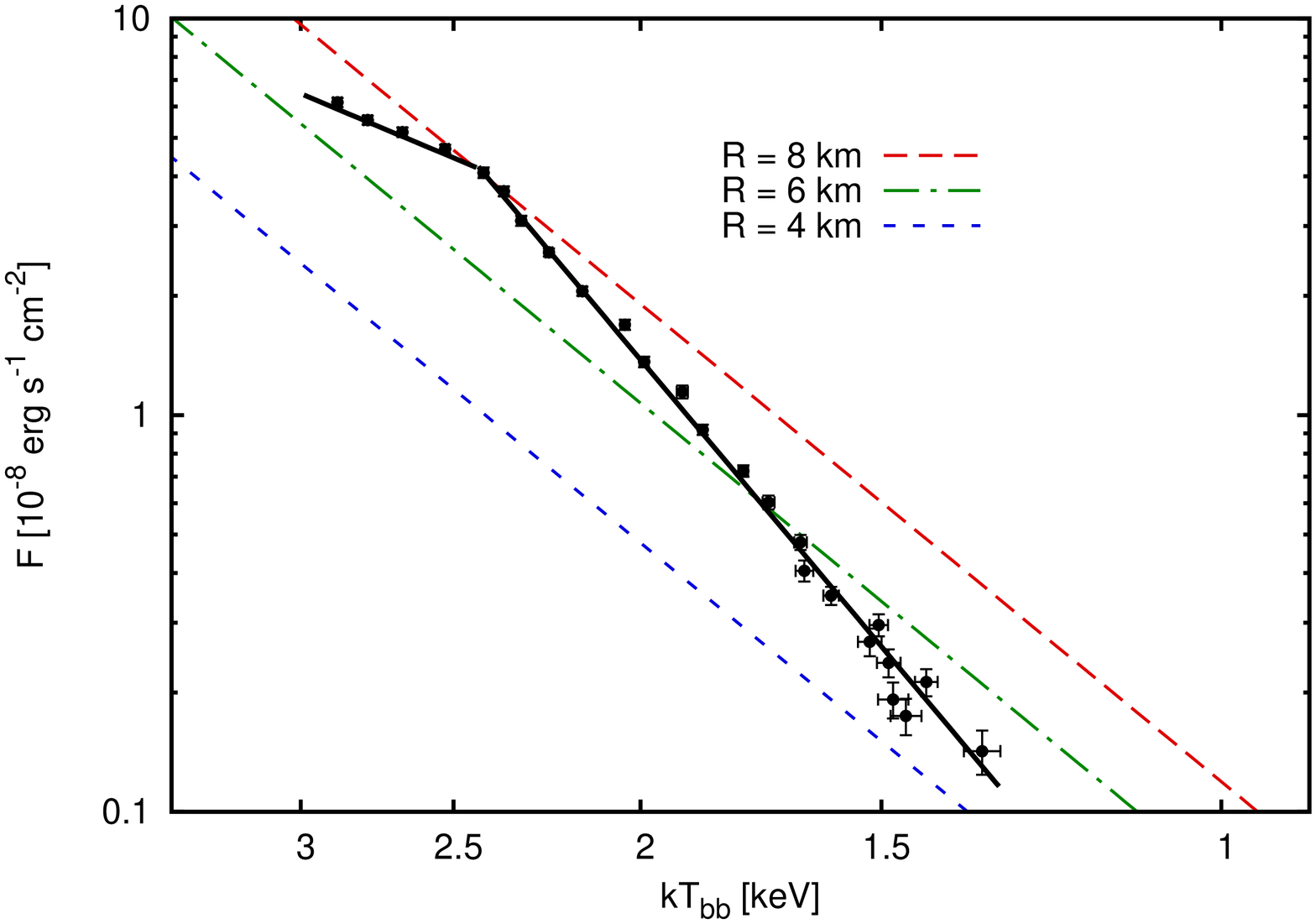}
\caption{Same data as in Figure~\ref{cooling}, rebinned into 25 points. The data are fitted by a broken power-law with indices $\nu_1=2.0\pm0.3$ and $\nu_2=5.72\pm0.06$ above and below the break, respectively. The break is at $kT_{\rm bb} = 2.43 \pm 0.02$~keV.} \label{cooling_rebined}
\end{figure}

\subsection{Flux-temperature relation during the cooling phase}

In Figure~\ref{cooling} we explore the relation between the bolometric flux, $F$, and the blackbody temperature, $kT_{\rm bb}$, during the cooling phase of all the busts. The diagonal lines in the plot represent lines of constant radius, $R=4$, 6 and 8~km, respectively, assuming a distance $d=7.6$~kpc to the source. Besides the fact that the dispersion is high at low flux levels, all bursts follow the same trend in this diagram (the spread in the plot is consistent with the statistical errors of the flux and temperature). Since the flux-temperature relation of all bursts is consistent with being the same, we binned the data sorted by flux into 25 points, and we calculated the weighted average of the flux and temperature within each bin. We also binned the data sorted by temperature, and we used different binning factors, but in all cases we obtained consistent results. In Figure~\ref{cooling_rebined} we plot the flux-temperature relation for the rebinned data. From this plot it is apparent that the cooling phase of the X-ray bursts in 4U 1820--30 does not follow the expected $F \propto (kT_{\rm bb})^4$ relation for an ideal blackbody of constant emitting area. Fitting the data with a single power law, $F \propto (kT_{\rm bb})^{\nu}$, gives $\nu$=4.95$\pm$0.03, but the fit is not good, with $\chi^2$=499 for 23 degrees of freedom. A broken power-law fits the data better (although formally it is not a good fit), with $\chi^2$=44 for 21 degrees of freedom, resulting in power-law indices before and after the break $\nu_1$=2.0$\pm$0.3 and $\nu_2$=5.72$\pm$0.06, respectively, and a break at $kT_{\rm bb}$=2.43$\pm$0.02~keV (see Figure~\ref{cooling_rebined}).

We also studied the flux-temperature relation dividing the bursts into two groups for ${\rm HC} > 1.05$ and ${\rm HC} < 1.05$, as in the previous subsection. After rebinning both sets of data into 25 points, we plotted their respective cooling phases and confirmed that both curves are consistent within errors, which lends support to the idea that the full sample of bursts in \mbox{4U 1820--30} is homogeneous.

\section{Discussion}

We present, for the first time, a homogeneous analysis of the cooling phase of all the thermonuclear X-ray bursts detected with {\it RXTE} in the ultracompact LMXB \mbox{4U 1820--30}. The sample contains 16 PRE bursts with a duration of 20$-$30~s, all of them detected when the source was in the hard state or in the transition from the hard to the soft state, at relatively low luminosity. 
The touchdown temperatures, bolometric peak fluxes, fluences and time-scales of the exponential cooling-tails are consistent with being the same in all bursts in the sample, supporting the idea that the bursts in 4U 1820--30 are very similar between them. Furthermore, contrary to what is expected for a constant-radius blackbody, the cooling phase of the X-ray bursts do not follow the $F \propto (kT_{\rm bb})^4$ relation. In fact, the relation between the bolometric flux and the blackbody temperature is fitted better by a broken power-law of indices 2.0$\pm$0.3 and 5.72$\pm$0.06 than with a single power law. The observed deviation from the $F \propto (kT_{\rm bb})^4$ relation complicates the use of the cooling phase of X-ray bursts to infer the mass and radius of the NS in \mbox{4U 1820--30}.

The global trend that we found for the cooling phase of the \mbox{X-ray} bursts in 4U 1820--30 is similar to that of the hard non-PRE bursts in 4U 1636--53 (Zhang et al. 2011). In both cases, the bolometric flux decays faster than what is expected for a blackbody with constant radius. This is reflected by the value of the power-law index ($\nu_2 = 5.72 \pm 0.06$) that best-fits the cooling phase of the bursts below $kT_{\rm bb} \sim 2.43$~keV (see Figure \ref{cooling_rebined}), and also in the best-fitting value for a single power law ($\nu = 4.95 \pm 0.03$) to the whole cooling-phase data. Taking into account the whole set of bursts in 4U 1820--30 observed with {\it RXTE}, there is no region of the $F$ vs. $kT_{\rm bb}$ diagram that can be fitted by a power-law with an index equal to 4. Our results are at variance with those of \cite{guver2010}, who analysed five bursts from 4U 1820--30 observed with {\it RXTE} (indicated with a $\dagger$ symbol in our Table~\ref{bursts_table}). \citet{guver2010} showed that during the cooling phase, three of these bursts followed the relation $F \propto (kT_{\rm bb})^4$, whereas they discarded the other two bursts because, according to their own explanation, these bursts showed significant and continuous fluctuations in the apparent radius. The three bursts analysed by \citet{guver2010} are included in our  Figure~\ref{cooling}, where it is apparent that none of the bursts in 4U 1820--30 follows the $F \propto (kT_{\rm bb})^4$ relation, even in the narrow flux range (1.0--6.0$\times10^{-8}$~erg~cm$^{-2}$~s$^{-1}$) that \citeauthor{guver2010} used. We fitted a power-law to these three bursts in this flux range obtaining an index of 4.8$\pm$0.16 (\mbox{1-$\sigma$} error; $\chi^2=15$ for 16 degrees of freedom), which is significantly different from 4. Fitting a power-law to the whole sample of bursts in the same flux range gives an index of 4.68$\pm$0.08 (\mbox{1-$\sigma$} error; $\chi^2=162$ for 98 degrees of freedom). We cannot explain the difference between our results and those of \citet{guver2010},  given that, as described in Section 2, we analysed the data as they did. We note, however, that from our analysis, all bursts in 4U 1820--30 observed with {\it RXTE} are consistent with following the same trend in the flux-temperature diagram during the cooling phase, including the three bursts studied by \citeauthor{guver2010}, but also those two bursts that they discarded.

The departure from the $F \propto (kT_{\rm bb})^4$ relation during the cooling phase of the X-ray bursts in 4U 1820--30 could be due to three different --but not mutually exclusive-- reasons: (i) the non-constancy of the true emitting area of the NS during this phase; (ii) changes of the colour-correction factor, $f_{\rm c}=T_{\rm bb}/T_{\rm eff}$, which accounts for hardening of the spectrum due to (Compton) electron scattering in the NS atmosphere \citep{london1986} during the cooling phase; (iii) the presence of a source of heat that increases the observed temperature, e.g. residual hydrogen nuclear burning during the cooling phase of the bursts. A final possibility is that the real underlying spectra are not blackbodies.

If $f_{\rm c}$ remains constant during the cooling phase, variations in $R_{\rm bb}$ could be explained by changes in the true emitting area of the NS. In that case, if a hotspot is formed away from the rotation pole, it should be possible to observe burst oscillations \citep[see e.g.][]{strohmayer2006}. However, no oscillations have been detected in the bursts of 4U 1820--30 \citep{galloway2008,watts2012}. Alternatively, if the emitting area does not change, variations in $f_{\rm c}$ could explain these differences. In terms of the spectral parameters, the colour-correction factor can be expressed as:

\begin{equation}
f_{\rm c}=\frac{T_{\rm bb}}{T_{\rm eff}}=\sqrt{\frac{R_{\infty}}{R_{\rm bb}}}=\sqrt{\frac{R_{\infty}}{d \sqrt{{\frac{F}{\sigma T_{\rm bb}^4}}}}}=\sqrt{\frac{R(1+z)}{d \sqrt{\frac{F k^4}{\sigma}}}}kT_{\rm bb} ,
\label{eq:fc}
\end{equation}

\noindent where $R_{\infty}$ is the NS radius seen by a distant observer, $z$ is the gravitational redshift at the surface of the NS, $d$ is the distance to the source, $R$ the radius of the NS, and $\sigma$ the Stefan-Boltzmann constant. 

\begin{figure}
	\centering
 \includegraphics[width=8.4cm]{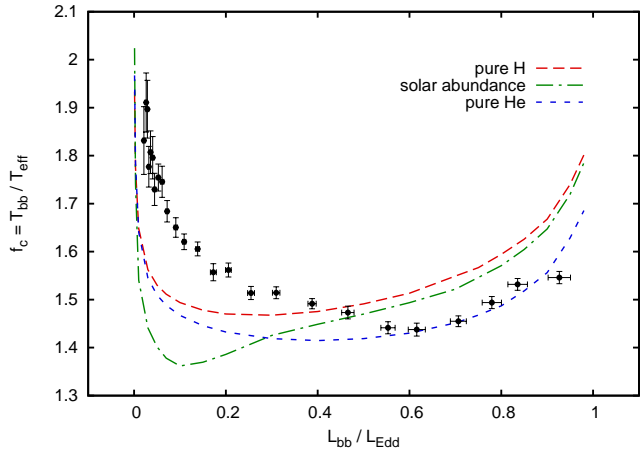}
\caption{Average colour-correction factor, $f_{\rm c}$, as a function of luminosity in Eddington units for the 16 X-ray bursts in \mbox{4U 1820--30}. The points correspond to the whole sample of bursts in 25 bins, using the data in Figure~\ref{cooling_rebined} and eq.~(\ref{eq:fc}). Dashed lines correspond to models of $f_{\rm c}$ for three different chemical compositions of the neutron-star atmosphere \citep{suleimanov2011}, indicated in the legend.}
\label{color_factor}
\end{figure}

In Figure~\ref{color_factor} we show the plot of the inferred $f_{\rm c}$ as a function of the luminosity in Eddington units. In the plot, we obtained the colour-correction factor from the spectral parameters, assuming $R=14$~km, $z=0.19$ and $d=7.6$~kpc. To estimate the luminosity in Eddington units we assumed that the flux reaches the Eddington limit at the touchdown point in each burst, and further that the emission is isotropic. Hence, we averaged the peak flux of the 16 bursts, obtaining $F_{\rm Edd}=6\times10^{-8}$~erg~cm$^{-2}$~s$^{-1}$. In the plot we also show the theoretical models of $f_{\rm c}$ from \cite{suleimanov2011} that correspond to atmospheres with three different chemical compositions: pure hydrogen, solar abundance and pure helium, respectively. Starting at near-Eddington luminosities, as the flux decreases, the inferred $f_{\rm c}$ slightly decreases until it reaches a minimum value $f_{\rm c}\sim$1.4 close to $L/L_{\rm Edd}=0.6$. After that, the inferred colour-correction factor starts to increase again, doing so even faster for $L/L_{\rm Edd}<0.2$, in the cooling tail of the bursts, close to the level of the persistent emission, where the errors in the fitted spectral parameters become important and the models are less accurate. Besides the fact that the models do not fit the data well, which prevents us from constraining the atmosphere composition, our results are incompatible with a solar-type atmosphere because the data do not show any dip at $L/L_{\rm Edd}<0.2$. At the beginning of the cooling, the data are consistent with models with He-rich composition, which agrees with the chemical composition of the accreted material \citep{rappaport1987} deduced from the properties of the white-dwarf companion. Moreover, the shape of $f_{\rm c}$ in this plot is similar to that of the hard non-PRE bursts in \mbox{4U 1636--53} (Zhang et al. 2011).

We note also that the distribution of $kT_{\rm bb}$ at constant bolometric flux levels broadens close to the tail of the bursts (Figure~\ref{cooling}). This could be due to differences in the underlying emission coming from the NS surface. This emission can be due to persistent accretion playing an important role when the burst emission ceases, or to the burning of a small residual fraction of hydrogen, not fully burned before touchdown, during the tail of the bursts. Hydrogen burning is slower than helium burning because the $\beta$-decay within the nuclear chain is driven by the weak force, thus having a longer time-scale than the triple-$\alpha$ process. The possibility that this source of heat operates during the cooling phase of the X-ray bursts \citep{cumming2003}, could explain the power-law index larger than four that best-fits the relation between the flux and the blackbody temperature.

\section*{Acknowledgements}

We are grateful to the anonymous referee for constructive comments that helped improve the paper. FG was supported by a fellowship from COSPAR and the Kapteyn Astronomical Institute. FG thanks the staff from the Kapteyn Astronomical Institute for their kind hospitality during his fruitful visit to the Institute. This research has made use of data obtained from the High Energy Astrophysics Science Archive Research Center (HEASARC), provided by NASA's Goddard Space Flight Center. This research made use of NASA's Astrophysics Data System.

\label{lastpage}
\end{document}